\documentclass[twocolumn,preprintnumbers,amsmath,amssymb,superscriptaddress]{revtex4-1}

\usepackage{graphicx,braket}
\usepackage{dcolumn}
\usepackage{bm}
\usepackage{color}

\begin{document}

\title{{Quantum-dot-like states in molybdenum disulfide nanostructures due to the interplay of local surface wrinkling, strain, and dielectric confinement}}


\author{Christian Carmesin}
\affiliation{Institute for Theoretical Physics, University of Bremen, P.O. Box 330440, 28334 Bremen, Germany}
\email{ccarmesin@itp.uni-bremen.de}
\author{Michael Lorke}
\affiliation{Institute for Theoretical Physics, University of Bremen, P.O. Box 330440, 28334 Bremen, Germany}
\affiliation{BCCMS, University of Bremen, P.O. Box 330440, 28334 Bremen, Germany}
\author{Matthias Florian}
\affiliation{Institute for Theoretical Physics, University of Bremen, P.O. Box 330440, 28334 Bremen, Germany}
\author{Daniel Erben}
\affiliation{Institute for Theoretical Physics, University of Bremen, P.O. Box 330440, 28334 Bremen, Germany}
\author{Alexander Schulz}
\affiliation{Institute for Theoretical Physics, University of Bremen, P.O. Box 330440, 28334 Bremen, Germany}
\affiliation{Present address: Leibniz Institute for Materials Engineering IWT, Badgasteiner Straße 3, 28359 Bremen, Germany}
\author{Tim O. Wehling}
\affiliation{Institute for Theoretical Physics, University of Bremen, P.O. Box 330440, 28334 Bremen, Germany}
\affiliation{BCCMS, University of Bremen, P.O. Box 330440, 28334 Bremen, Germany}
\author{Frank Jahnke}
\affiliation{Institute for Theoretical Physics, University of Bremen, P.O. Box 330440, 28334 Bremen, Germany}


    \begin{abstract}
The observation of quantum light emission from atomically thin transition metal dichalcogenides has
opened a new field of applications for these material systems.
The corresponding excited charge-carrier localization has been linked to defects and strain, while open questions remain regarding the microscopic origin.
We demonstrate that the bending rigidity of these materials leads to wrinkling of the two-dimensional layer.
The resulting strain field facilitates strong carrier localization due to its pronounced influence on the band gap.
Additionally, we consider charge carrier confinement due to local changes of the dielectric environment and show that both effects contribute to modified electronic states and optical properties.
The interplay of surface wrinkling, strain-induced confinement, and local changes of the dielectric environment is demonstrated for the example of 
nanobubbles that form when monolayers are deposited on substrates or other two-dimensional materials.
   \end{abstract}

    \maketitle
    
In the past, single-photon emission has been realized using trapped-ion systems~\cite{ion}, 
NV centers in diamond~\cite{PhysRevLett.100.077401}, and epitaxially grown quantum dots~\cite{QD} with important applications 
in quantum information. 
Recently, single-photon emission from spatially localized centers in transition metal 
dichalcogenides (TMDs) has been 
demonstrated~\cite{srivastava_optically_2015, koperski_single_2015, chakraborty_voltage-controlled_2015, he_single_2015, kern_nanoscale_2016, branny_discrete_2016,kumar_strain-induced_2015,branny_deterministic_2017, palacios-berraquero_large-scale_2017,khestanova_universal_2016,shepard_nanobubble_2017}.
For this purpose, TMD monolayers have been placed over gold 
nanostructures~\cite{kern_nanoscale_2016, branny_discrete_2016}, substrates with etched holes~\cite{kumar_strain-induced_2015}, and 
arrays of dielectric micropillars~\cite{branny_deterministic_2017, palacios-berraquero_large-scale_2017}. 
In the latter case, up to 96\% active single-photon emitters have been obtained. 
Quantum light emission has been also achieved with nanobubbles, which are formed in vertically stacked TMD structures~\cite{khestanova_universal_2016, shepard_nanobubble_2017}.

Single-photon emission from atomically thin TMDs can originate either from structural defects resulting in 
electronic trap states, which are known to be a source of photoluminescence centers~\cite{Tongay:13,ChowP:15}, or due to 
strain-induced potentials with three-dimensional carrier confinements. 
In this paper, we show that the strain-induced occurrence of localized electronic states is intrinsically linked to a 
non-continuous deformation of the TMD surface. 
We find that bending of the atomically thin sheet leads to formation of local surface wrinkles and bond deformations are identified as their microscopic origin. 
Due to the strong dependence of the TMD band gap on local strain fields~\cite{Steinhoff:14} 
this wrinkling induces a spatially localized carrier confinement. 
We show that this wrinkling takes place on length scales of $1\,$nm, providing efficient three-dimensional localized states and thereby enabling single-photon emission. 
This supports recent experimental results~\cite{branny_discrete_2016}, which first established the connection between surface wrinkling and localized emission, 
although the resolution of the used confocal microscopy is limited to much larger $\mu$m length scales. 
The local wrinkling effect is different from surface corrugations on a $10-20\,$nm scale, that is known for many 2D materials, as studied for graphene in Ref.~\cite{Meyer:07}.


%
\begin{figure*}[!ht]
\includegraphics[width=\textwidth]{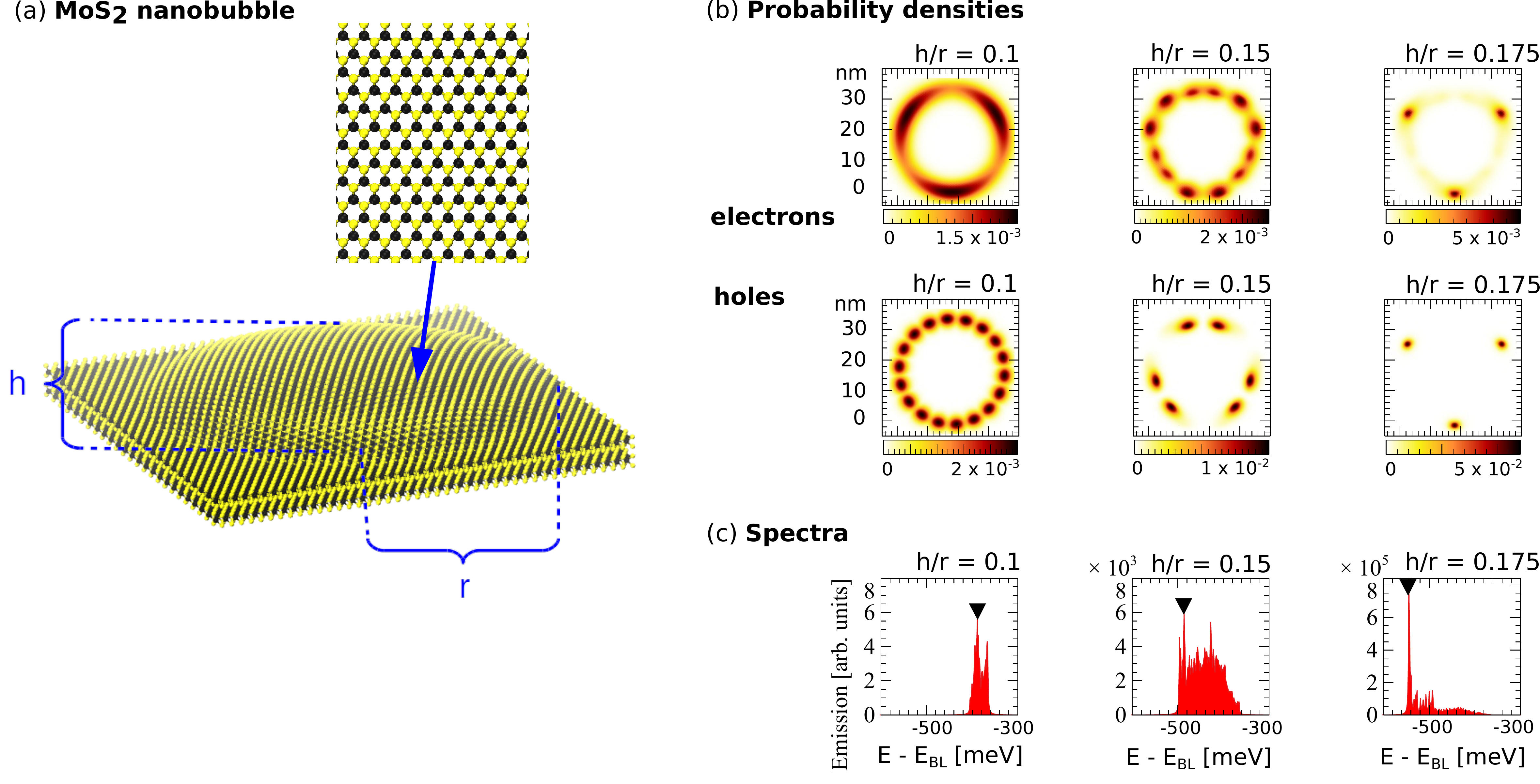}
\caption{(a) MoS$_2$ air nanobubble with radius $r$ and height $h$.
(b) Top view of the probability densities for electrons and holes, which are responsible for the strongest optical transition marked in (c). Results are shown for a bubble radius of $r=18\,$nm and various aspect ratios $h/r$. 
(c) Emission spectra calculated from Fermi's golden rule based on the tight-binding states for the aspect ratios in (b).
 \label{Abb1}}%
\end{figure*}

To investigate the interplay between TMD layer deformations and carrier confinement, we model a TMD layer with atomic resolution using a million-atom supercell. Based on valence force field simulations, new equilibrium positions of the individual atoms in the bended material are determined. The information about the displaced atoms is used in a tight-binding electronic-state calculation for the supercell structure. For the example of nanobubbles, we demonstrate that surface wrinkling acts as the main cause for quantum confinement with quantum-dot-like electronic states. The considered geometry is depicted in Fig.~\ref{Abb1}(a). Nanobubbles are formed when placing a monolayer of MoS$_2$ on another MoS$_2$ sheet used as a flat substrate, which is comparable to recent experimental investigations~\cite{khestanova_universal_2016}. Starting from a paraboloid profile of the nanobubble, the relaxation of atoms in the upper TMD layer is performed while keeping the atoms in the lower layer fixed. For this part, we utilize a REAX potential \cite{ReaxFF} with the parametrization from Ref.~\cite{ostadhossein_reaxff_2017} within a valence force field calculation, which is capable of accurately describing bond deformations under strain as well as continuous bond formation and breaking dynamics. We find that both in-plane strain and bending contribute to the shape of the nanobubble (see Supporting Information for more details). Subsequent electronic-state calculations use this information about the positions of the individual atoms. For a supercell with an in-plane extension of 130$\,$nm and up to 1.2$\times10^6$ atoms, a 6-band tight-binding Hamiltonian is solved using the parametrization given in the Supporting Information. 
Strain-induced local band gap changes arising from the displaced atomic positions are included via a generalized Harrison rule~\cite{froyen_elementary_1979}. 
Additionally, when locally detaching the upper layer from the substrate underneath and changing from a commensurate bilayer to monolayer-like structures across the nanobubble, a modified dielectric environment and electronic hybridization is expected~\cite{cheiwchanchamnangij_quasiparticle_2012, rosner_two-dimensional_2016, raja_coulomb_2017, dielectric_nanoletter_florian}. We include both effects into our model by changing individual tight-binding parameters based on GW calculations, as explained in the Supporting Information. Based on the calculated tight-binding states, optical emission spectra are obtained using Fermi’s golden rule.

In our analysis, we consider nanobubbles of different aspect ratios $h/r$ and sizes.
The degree of localization is controlled by the aspect ratio, as shown in Fig.~\ref{Abb1}(b).
In contrast, the nanobubble size does not change the qualitative behavior, as presented in the appendix.
For $h/r=0.1$ the single-particle states are delocalized along the edge of the bubble, whereas for an increasing 
aspect ratio of $h/r=0.15$ distinct maxima are found.
Due to the increased strain, these maxima carry the $C_{3v}$ symmetry of the underlying crystal lattice.
The trend of stronger localization with increasing aspect ratio continues for $h/r=0.175$, which corresponds to 
the experimentally reported value of Ref~\cite{khestanova_universal_2016}.
In this case, the probability densities are localized at three distinct positions at the edge of the nanobubble. 
%

%
Figure~\ref{Abb1}(c) provides the emission spectra corresponding to the three different aspect 
ratios in Fig.~\ref{Abb1}(b), obtained from all confined states of the TB model and Fermi's golden rule.
For $h/r=0.1$, the oscillator strength is very weak due to a small dipole interaction matrix element, which is the result of 
a much larger radial broadening of the probability density for the electrons than 
for the holes.
More optical transitions are present for larger aspect ratios $h/r$.
In the case of $h/r=0.175$ strong emission from the electron and hole states, shown in the right panel of Fig.~\ref{Abb1}(b), 
as well as a broad background emission from other states is obtained.
A similar behavior has been observed in recent experiments with WSe$_2$ nanobubbles~\cite{shepard_nanobubble_2017} and can 
be explained by increasing strain, which leads to a deeper confinement, stronger localization of the single-particle states,
as seen in Fig.~\ref{Abb1}(b), and a redshift of the optical spectra.

%
\begin{figure*}[!ht]
\includegraphics[width=\textwidth]{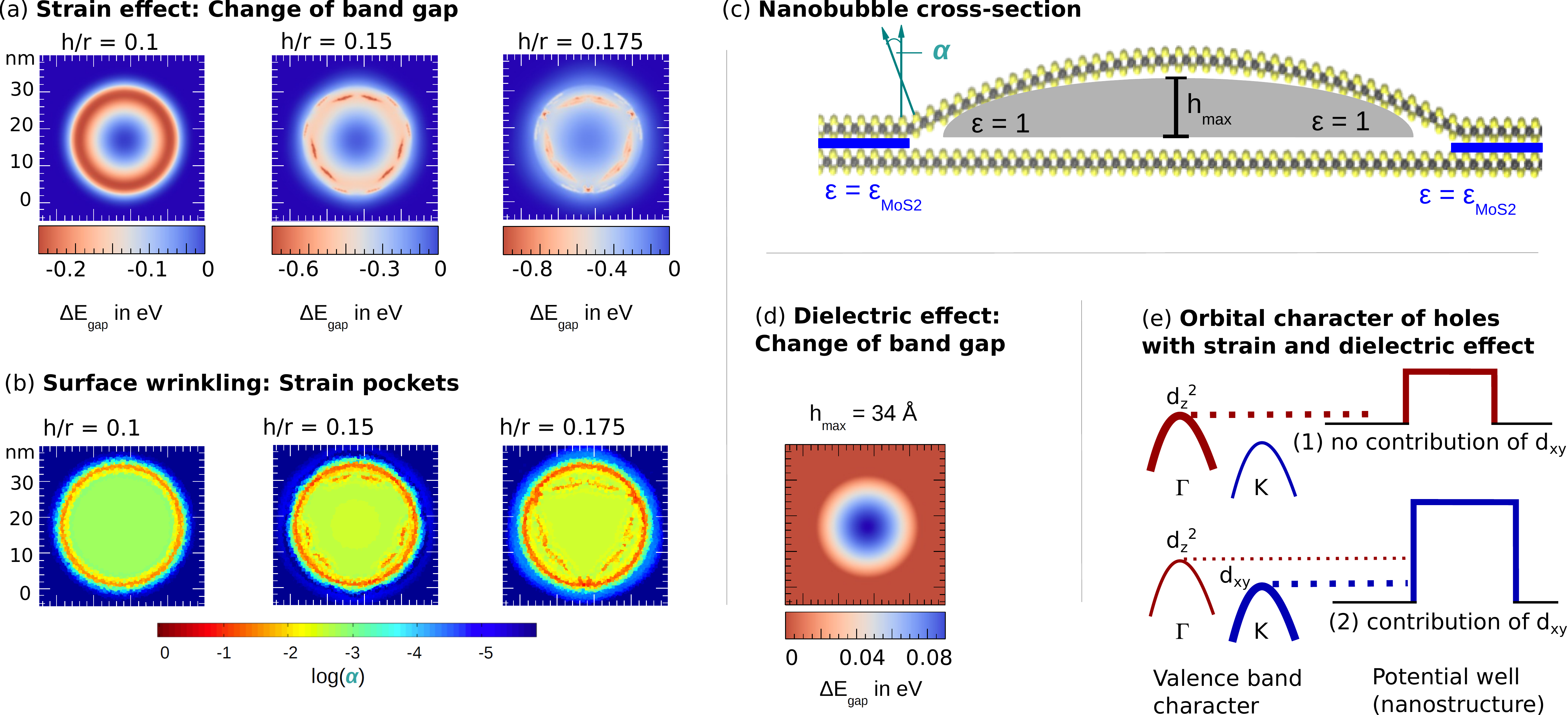}
\caption{(a) Strain-induced band gap shifts for different aspect ratios $h/r$.
(b) Average angle $\alpha$ (in radians) between normal vectors of neighboring unit cells reveal wrinkling that corresponds to the spatially localized strain pockets. 
(c) Cross-section through the nanobubble of height $h_{\text{max}}$. The geometric situation implies a change of the dielectric and electronic environment for the MoS$_2$ layer that is forming the nanobubble.
The arrows indicating an angle $\alpha$ between normal vectors of neighboring unit cells on the discretized surface as used in (b).
(d) Repulsive potential inside the nanobubble of height $h_{\text{max}} = 34\,\text{\AA}$ and aspect ratio $h/r = 0.175$ as a result of the modified dielectric and electronic environment for the top layer. 
(e) Schematic representation of the orbital character of localized hole states. For shallow confinement, the states are composed of $d_z$ orbitals, while for deeper confinement $d_{xy}$ orbitals also contribute. 
 \label{Abb2}}%
\end{figure*}
To analyze the mechanism behind the strong carrier localization, we quantify the contributions 
to the confinement potential due to strain as well as local changes of the screening and electronic 
hybridization for the upper TMD layer.
As a result of the nanobubble geometry, shown in Fig.~\ref{Abb2}(c), the dielectric and electronic
environment outside of the bubble corresponds to a bilayer, while, in the center of the bubble,  
the large distance between top and bottom layer resembles the situation  of a monolayer.
This effect increases with increasing height $h_{\text{max}}$ of the bubble, as illustrated in Fig.~\ref{Abb2}(c). 
The resulting changes of band gap are depicted in Fig.~\ref{Abb2}(d).
Effectively, it leads to a repulsive potential in the center of the bubble.
From our calculations we obtain band gap changes on the order of $100\,$meV, consistent with recent measurements~\cite{raja_coulomb_2017}.

The strain-induced changes of band gap for three different aspect ratios are displayed in Fig.~\ref{Abb2}(a).
By comparing these strain effects with the probability densities in Fig.~\ref{Abb1}(b), we can conclude that the strain-induced changes 
of the band gap are responsible for the increasing localization, as the dielectric effect is rotationally symmetric (see Fig.~\ref{Abb2}(d)).
For aspect ratios of $h/r=0.15$ and $h/r=0.175$, strain pockets are found.
To illuminate their origin, the average angle between normal vectors at neighboring unit cells (schematically 
depicted in Fig.~\ref{Abb2}(c)) is displayed in Fig.~\ref{Abb2}(b), as described by the normal-normal correlation function (see Supporting Information).
These average angles reveal a local wrinkling of the surface for aspect ratios of $h/r=0.15$ and $h/r=0.175$, 
with the angle of the normal vectors between neighboring unit cell varying up to $6^{\circ}$.
A comparison between Figs.~\ref{Abb2}(a) and (b) shows that the strain becomes non-uniform and peaks in the wrinkles.
The physical origin of this behavior can be traced back to the high Young's modulus of MoS$_2$ 
(see e.g. Ref.~\cite{ostadhossein_reaxff_2017, TMD-mechanics}).
The higher the modulus, the better a material resists against compression/elongation.
Especially for a high modulus, the material tends to minimize its total change of length. In particular, this is possible by the formation of wrinkles.
The bond deformations responsible for the wrinkling are more prominent in TMDs than in graphene, as the 
former consist of a tri-atomic structure, in which the transition metal atoms are sandwiched between chalcogen atoms.
Additionally, the spherical deformation of the material that is imposed by the nanobubble geometry favors wrinkle formation. 
We expect the wrinkles to be a general feature of spherical deformations,
in contrast to depositing the material on a cylindrical surface or over a ridge. 
Mathematically, the difference between these 
situations lies in the Gaussian curvature~\cite{Nakahara} of the geometry that is imposed 
upon the TMD monolayer.
The strong band gap variations, induced by strain and the dielectric environment, also
influence the oscillator strength of the optical transitions in Fig.~\ref{Abb1}(c). The main reason for this change is illustrated in Fig.~\ref{Abb2}(e). 
The localized electron states, originating from the conduction band, inherit the $d_{z^2}$ orbital character of the $\mathbf{K}$-point. 
In the strained bubble geometry, the $\mathbf{\Gamma}$ valley lies energetically higher than the $\mathbf{K}$-valley for holes.
Depending on the depth of the confinement potential, the orbital character of the localized hole states is
than either determined by states originating from the $\mathbf{\Gamma}$-point ($d_{z^2}$), or from both the $\mathbf{\Gamma}$- and the $\mathbf{K}$-point ($d_{xy}$). 
In the former case, the emission is weak, as the electron and hole wave functions carry the same orbital character ($d_{z^2}$). In the latter case, the strong dipole matrix
element of the $\mathbf{K}$-point transition dominates, as known from the 2D monolayer~\cite{Steinhoff:14}.
These large differences in the dipole matrix elements explain why the measured recombination times of a few nanoseconds~\cite{srivastava_optically_2015, shepard_nanobubble_2017} for localized states in TMD nanostructures are much longer than the sub-picosecond range, which is observed for the 2D monolayer~\cite{poellmann_radiative_decay2015}.

In summary, our calculations provide insight into the mechanism of quantum-dot-like carrier localization in TMD nanostructures for the example of MoS$_2$ nanobubbles.
We find a strong localization on the nanometer scale inside strain pockets, which carry the $C_{3v}$ symmetry of the underlying crystal lattice. 
It is demonstrated that the formation of such strain pockets is caused by a non-uniform strain, which peaks in surface wrinkles, an effect that is inaccessible to continuum theory. 
Our method is based on the requirements of covering strain as well as local changes of the screening and electronic hybridization on the nanometer scale. 
We quantify the contributions of these effects to the confinement potential in a combined atomistic valence force-field and million-atom supercell tight-binding approach.
Regarding the emission properties, it is shown that the orbital character of the localized states changes compared to the 2D monolayer, 
explaining the measured recombination times of a few nanoseconds~\cite{srivastava_optically_2015, shepard_nanobubble_2017} in TMD nanostructures.
We expect these properties and mechanisms to be present in other TMD nanostructures, since the underlying physical properties are universal for TMD materials.
Our results are expected to stimulate new and alternative realizations of strong carrier confinement in atomically thin TMD semiconductors.

\section*{Acknowledgements}
The authors gratefully acknowledge computational resources from HLRN Berlin and funding from the DFG ("Deutsche Forschungsgemeinschaft") via the graduate school "Quantum-Mechanical Material Modeling".

\onecolumngrid
\appendix*

\section{Nanobubble geometry}
Starting point of our relaxation procedure is a rectangular supercell of a  MoS$_2$ bilayer with periodic boundary conditions and an interlayer equilibrium distance of $3.0\,\text{\AA}$ in accordance with~Ref.~\cite{ostadhossein_reaxff_2017}.
The top layer, which subsequently forms the bubble, is shorter in both lateral directions so that its atoms are not influenced by the boundary conditions.
To generate the bubble, the atoms of the top layer are displaced onto a paraboloid with a given height, corresponding to the desired aspect ratio.

In order to determine the relaxed atomic positions, we employ the LAMMPS code~\cite{plimpton_fast_1995} using the REAX potential of Ref.~\cite{ostadhossein_reaxff_2017}.
To achieve a convergent geometry, we utilize a supercell of dimensions $130\,$nm$\,\times\,130\,$nm$\,\times\,15\,$nm, containing about $1.2\cdot 10^6$ atoms.
The relaxation is performed until a force tolerance of $10^{-3}\,\text{eV}/\text{\AA}$ or a relative energy tolerance of $10^{-16}$ is reached.
In each iteration step, the charge equilibration energy is minimized until it reaches a value below $10^{-3}\,$eV.
\begin{figure}[ht!]
	\centering
	\includegraphics[width=0.99\linewidth]{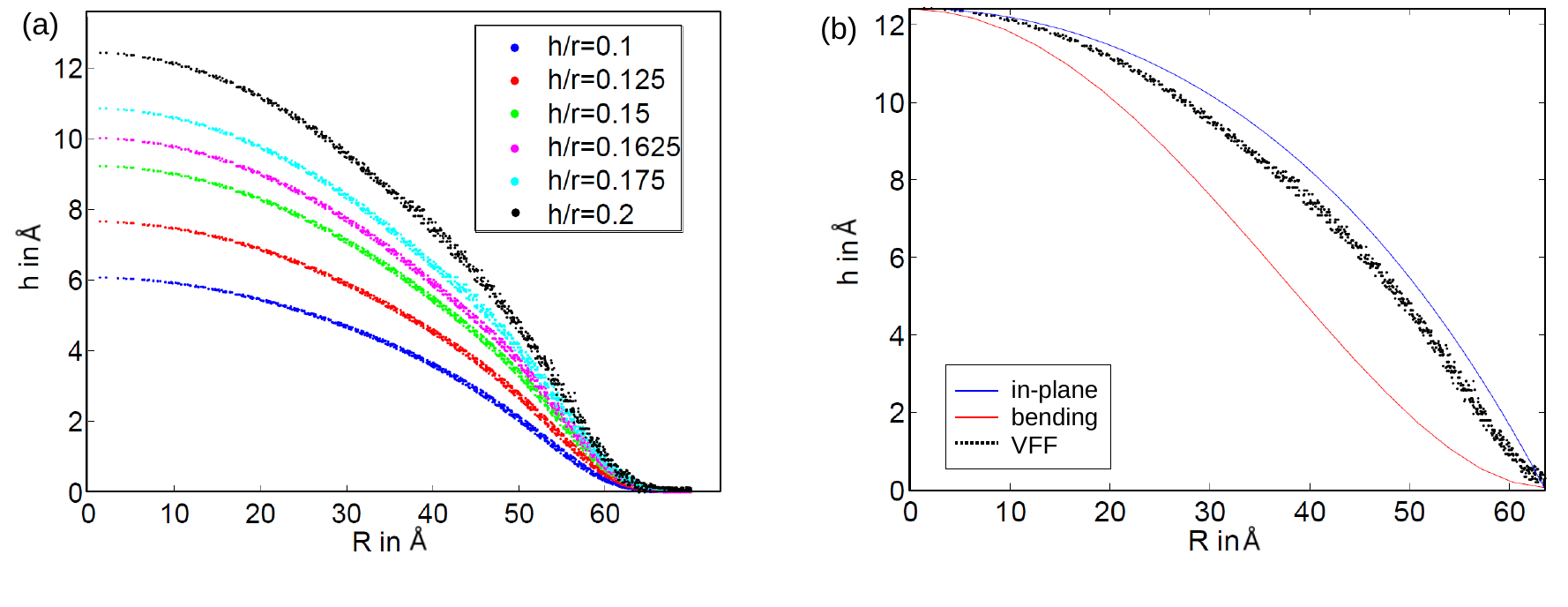}
	\caption{ Height profiles of MoS$_2$ bubbles. (a) Curves for different aspect ratios.
	(b) Valence force field results for an aspect ratio of $h/r = 0.2$ and continuum elasticity results from E. Khestanova et al.~\cite{khestanova_universal_2016} for dominantly in-plane strains (blue line) and bending (red line). }
	\label{profile}
\end{figure}
The obtained atomic positions provide access to height profiles of MoS$_2$ bubbles for different aspect ratios $h/r$, which are depicted in Fig.~\ref{profile}(a).
In Fig.~\ref{profile}(b), these values are compared to continuum elasticity theory \cite{khestanova_universal_2016}, 
where two cases, dominated by in-plane strain and bending, can be distinguished.
Since our microscopically calculated height profile of the bubble is in between these limits, we conclude that both aspects contribute.

\subsection{Monolayer tight-binding model}
\begin{figure}[ht!]
	\centering
	\includegraphics[width=0.85\linewidth]{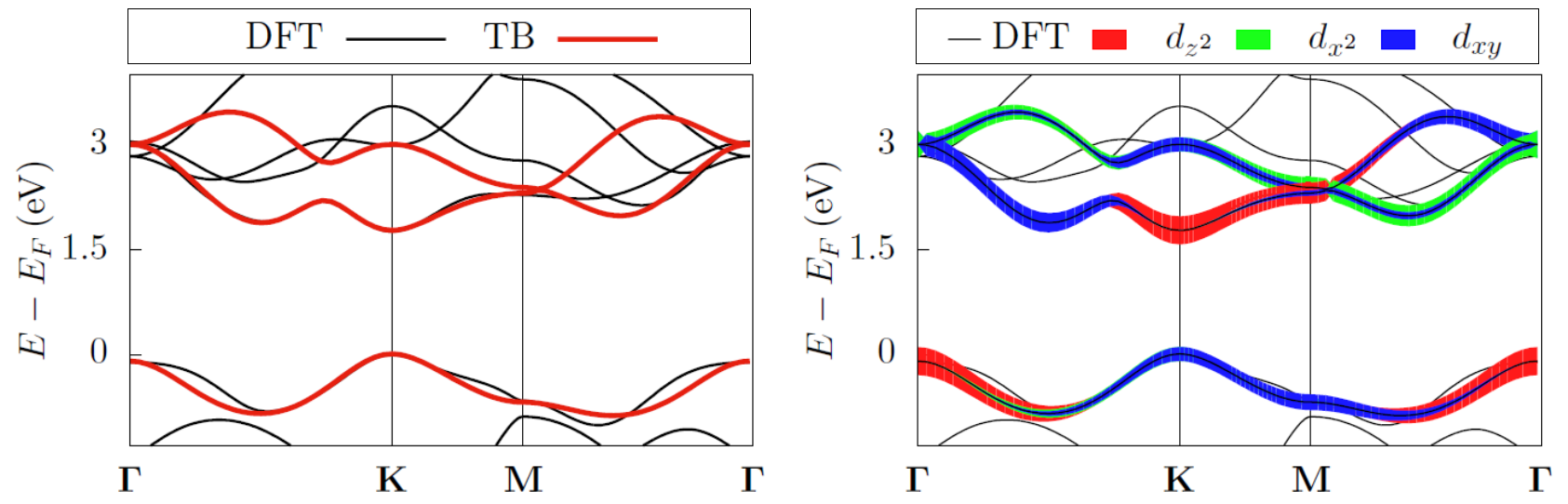}
	\caption{Tight-binding band structure. Left: Comparison of the three-band monolayer MoS$_2$ TB band structure (red) with a DFT calculation (black). Right: Character of the three bands $d_{z^2}$ (red), $d_{x^2}$ (green) and $d_{xy}$ (blue).}
\label{MoS2bands}
\end{figure}

To describe the TMD monolayer, we employ an effective TB Hamiltonian with nearest and 
next-nearest neighbor interactions between Mo atoms and a $\{d_0, d_{+2}, d_{-2}\}$ basis set.
The respective Hamilton matrix $H$ in $\mathbf{k}$-space is given by
\begin{equation}
H=
  \begin{pmatrix}
    H_{d_0d_0} & H_{d_0d_2} & H_{d_0d_{-2}} \\
    H^{*}_{d_0d_2} & H_{d_2d_2} & H_{d_2d_{-2}} \\
    H^{*}_{d_0d_{-2}} & H^{*}_{d_2d_{-2}} & H_{d_{-2}d_{-2}}
  \end{pmatrix},
\end{equation}
with matrix elements
\begin{equation}
H_{\alpha\beta} = \braket{\alpha,k|H|\beta,k} = \sum_{\mathbf{R}} e^{i\mathbf{k}\cdot\mathbf{R}}\braket{\alpha,0|H|\beta,\mathbf{R}}~.
\end{equation}
For the atomic positions $\mathbf{R}$, we consider nearest neighbors $\mathbf{R}_1,..,\mathbf{R}_6$
and next nearest neighbors $\mathbf{R}_7,..,\mathbf{R}_{12}$,
using the in-plane coordinates
\begin{equation}
\begin{split}
&\mathbf{R}_1 = (1,0), \,\,\,\,\, \mathbf{R}_2 = (1,1), \,\,\,\,\,
\mathbf{R}_3 = (0,1), \\
&\mathbf{R}_4 = (-1,0), \mathbf{R}_5, = (-1,-1) \,\,\,\,\,
\mathbf{R}_6 = (0,-1), \\ \, 
&\mathbf{R}_7 = (2,1), \,\,\,\,\, \mathbf{R}_8 = (1,2), \,\,\,\,\,
\mathbf{R}_9 = (-1,1), \\
&\mathbf{R}_{10} = (-2,-1), \,\,\,\,\,        \mathbf{R}_{11} = (-1,-2) \,\,\,\,\, \mathbf{R}_{12} = (1,-1)~
\end{split}
\end{equation}
in units of the the Bravais-lattice vectors
\begin{equation}
\mathbf{a}_1 = a\begin{pmatrix} 1 \\ 0 \end{pmatrix}
~~~~~~~~~~~~~~~
\mathbf{a}_2 = a\begin{pmatrix} -1/2 \\ \sqrt{3}/2 \end{pmatrix}
\end{equation}
with the lattice constant $a = 3.18\,\text{\AA}$.
The matrix elements
\begin{equation}
\braket{\alpha,0|H|\beta,\mathbf{R}} = f_{\alpha,\beta}(\phi)t_{\alpha\beta}(|\mathbf{R}|)
\end{equation}
include the TB hopping parameter $t_{\alpha\beta}(R)$ towards nearest and next-nearest neighbors of the Mo-atoms, respectively.
\begin{table}[t]
\centering
\begin{tabular}{l c}
\hline
\hline
\vspace{2mm}
$\varepsilon_{0}$ & -2.17187\,eV \\\vspace{2mm}
$\varepsilon_{2}$ & -2.07972\,eV \\\vspace{2mm}
$t_{0}$  & -0.326583\,eV \\\vspace{2mm}
$t_{0\pm2}$ & 0.561734\,eV\\\vspace{2mm}
$t_{2}$ & -0.411327\,eV\\\vspace{2mm}
$t_{\pm2 \mp2}$ & -0.355268\,eV\\\vspace{2mm}
$t'_{0}$  & -0.0537226\,eV \\\vspace{2mm}
$t'_{0 \pm2}$ & -0.00459159\,eV\\\vspace{2mm}
$t'_{2}$ & 0.052774\,eV\\\vspace{2mm}
$t'_{\pm2 \mp2}$ & -0.123627\,eV\\
\hline
\hline
\end{tabular}
\caption{TB parameters for the MoS$_2$ monolayer. Orbital energies and coupling matrix elements are denoted by $\varepsilon$ and $t$, respectively. The primed terms describe next-nearest neighbor couplings.}
\label{MoS2-parameter}
\end{table}
While the threefold symmetry of the MoS$_2$ lattice would be enhanced to a six-fold symmetry when considering the Mo sublattice only, 
we retain the original C$_{3v}$ symmetry by including different phase factors $\phi$ towards nearest and next-nearest neighbours
\begin{equation}
f_{\alpha,\beta}(\phi)=
\begin{cases}
\text{nearest neighbors:} \\
\text{exp}(-i\alpha(\phi-\pi/6)+i\beta(\pi+\phi+\pi/6)),~&\text{if}~\phi =0,2\pi/3,4\pi/3\\
\text{exp}(-i\alpha(\phi+\pi/6)+i\beta(\pi+\phi-\pi/6)),~&\text{if}~\phi =\pi/3, \pi, 5\pi/3\\
\text{next nearest neighbors:} \\
\text{exp}(-i\alpha\phi+i\beta(\pi+\phi)),~&\text{if}~\phi =\pi/6, 3\pi/2, 5\pi/6 \\
\text{exp}(-i\alpha(\pi+\phi)+i\beta \phi),~&\text{if}~\phi =\pi/2, 7\pi/6, 11\pi/6.
\end{cases}
\end{equation}
Following Ref.~\cite{PhysRevB.88.085433}, we include spin-orbit coupling according to
%
\begin{gather}\label{HMoSO}
H^{SO}=
 \left[\begin{array}{cc}
 H + \frac{\lambda}{2} L_z & 0  \\
0 & H - \frac{\lambda}{2} L_z
\end{array}\right]
\end{gather}
with
\begin{gather}
L_z=
 \left[\begin{array}{ccc}
 0 & 0 & 0  \\
0 & 0 & 2i  \\
0 & -2i & 0
\end{array}\right].
\end{gather}
The Hamiltonian in Eq.~\eqref{HMoSO} is block diagonal, which means
that the $z$-component of the spin is not mixed by the spin-orbit coupling and serves as a good quantum number due to the $\sigma_h$ symmetry~\cite{PhysRevB.88.085433}.
By using our TB approach, the band structure of DFT calculations is well approximated, as shown in Fig.~\ref{MoS2bands}. 
The DFT calculations were performed using the VASP code~\cite{Kresse1, Kresse2} together with the supplied projector augmented wave (PAW) potentials~\cite{Kresse3} and a 16 x 16 x 1 $\Gamma$-point sampling of the Brillouin zone.
The used TB parameters for a MoS$_2$ monolayer are given in Table.~\ref{MoS2-parameter}.

\subsubsection*{Hamiltonian at points of high symmetry}
Both strain and changes of the dielectric environment influence the band gap of MoS$_2$.
To incorporate these effects into the TB model, we diagonalize the TB Hamiltonian analytically at points of high symmetry, 
which allows for an identification of the TB parameters to be adapted.
For the $\mathbf{\Gamma}$-point we find
\begin{equation}
H(\mathbf{\Gamma}) = \begin{pmatrix}
\varepsilon_0 +6t_0 + 6t'_0 & 0 & 0  \\
0 & \varepsilon_2 - 3t_2 + 6 t'_2 & 0 \\
0 & 0 & \varepsilon_2 - 3t_2 + 6 t'_2  \\
\end{pmatrix}
\label{Gamma}
\end{equation} 
with the first diagonal element, denoting the eigenvalue of the valence band (VB) and the other two, describing the conduction bands (CBs).
At the $\mathbf{K}$-point, we obtain
\begin{equation}
H(\mathbf{K}) = \begin{pmatrix}
\varepsilon_2 +6 t_2  +6t'_2		& 0 								& 0  \\
0								& \varepsilon_0 -3t_0 +6t'_0  	& 0   \\
0 								& 0								& \varepsilon_2 -3 t_2 +6t'_2 \\
\end{pmatrix}.
\label{K}
\end{equation}
In the next section, we discuss how these results can be used to incorporate the effects of strain and of the dielectric environment into the TB Hamiltonian.

\subsection{Tight-binding model for nanobubbles}

\subsubsection*{Implementation of strain}
\begin{figure}[ht!]
	\centering
	\includegraphics[width=0.45\linewidth]{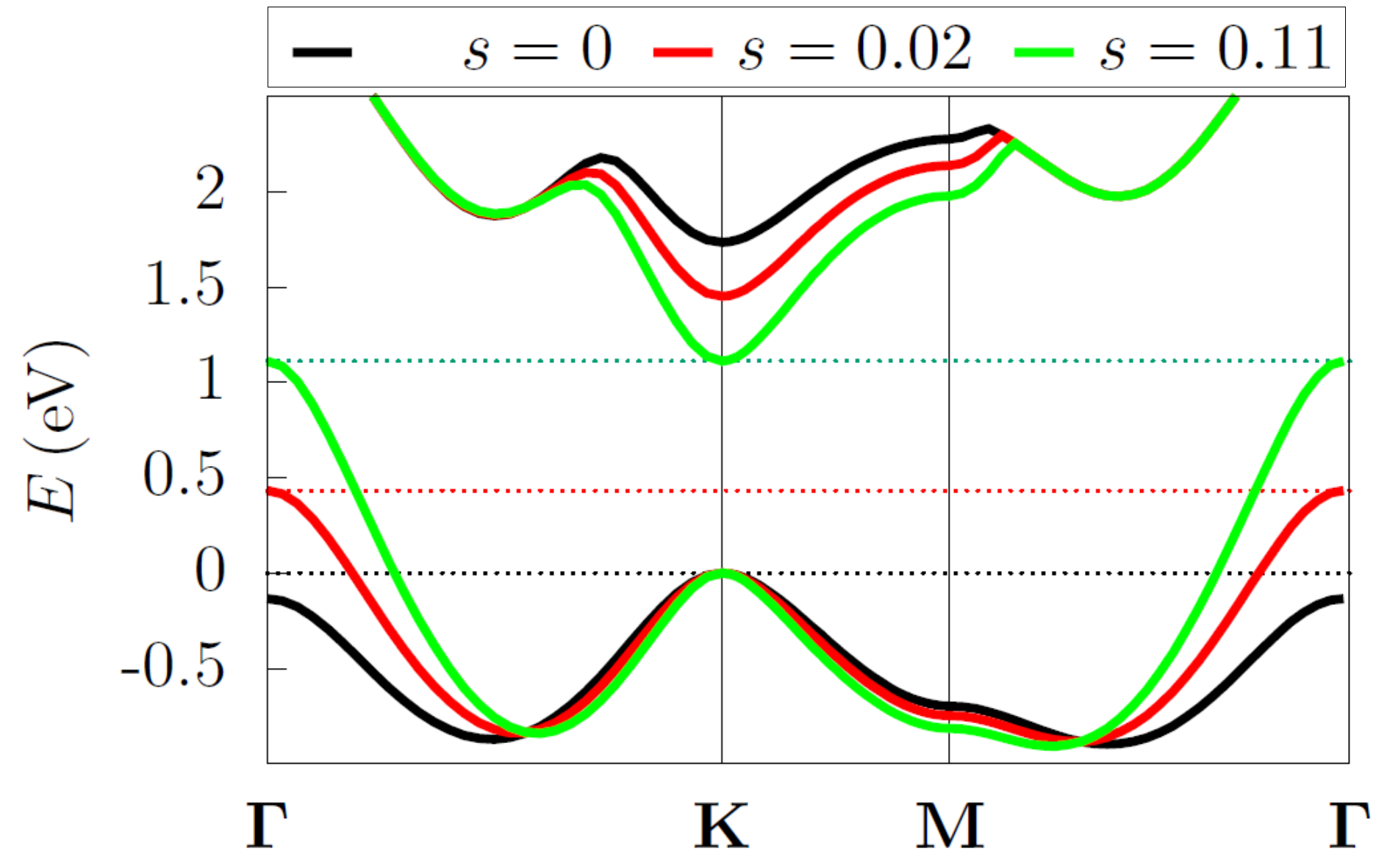}
	\caption{
	Band structure of a MoS$_2$ monolayer for different strain situations $s=(r- a)/a$. 
	The dashed lines indicate Fermi levels. }
	\label{bandgap}
\end{figure}

In the case of a MoS$_2$ monolayer, strain leads to an transition between a direct band gap at the $\mathbf{K}$-point and an indirect band gap 
between the $\mathbf{\Gamma}$-point of the VB and the $\mathbf{K}$-point of the CB, as demonstrated for instance in Ref.~\cite{PhysRevB.87.235434}.
From Eqs.~\eqref{Gamma} and \eqref{K}, we obtain 
\begin{equation}
E_{\text{CB}}(\mathbf{K}) -E_{\text{VB}}(\mathbf{\Gamma}) =-9t_0.
\end{equation}
Hence, we can modify the nearest neighbor hopping parameter $t_0$ to describe the direct to indirect transition in strained monolayers.
%
We employ a 
generalized Harrison rule~\cite{froyen_elementary_1979} 
\begin{equation}
t_0(r) = t_0 /\left(1+s \right)^{\eta}
\end{equation}
with $s=(r-a)/a$, where $r$ is the inter-atomic distance and $a$ the lattice constant.
For an exponent of $\eta = 11$ the experimental band gaps in Ref.~\cite{conley_bandgap_2013} are reproduced.
The band structure for different strain situations within this approach is depicted in Fig.~\ref{bandgap}.
\subsubsection*{Implementation of the dielectric environment}

In addition to strain, the band gap can also be modified by changing the dielectric environment of the layer \cite{rosner_two-dimensional_2016,raja_coulomb_2017}.
While the system resembles a monolayer in the middle of the bubble, it corresponds to a bilayer at the edges.
In this transition, not only the value of the band gap changes, but also its character.
ARPES measurements \cite{PhysRevLett.111.106801} demonstrate
that the MoS$_2$ monolayer band gap is direct, whereas the bilayer band gap is indirect.
Additionally, the value of the band-gap at the $\mathbf{K}$-point is reduced, due to more efficient dielectric screening.
To determine this band gap $E^g_{\mathbf{K}}(z)$ as a function of the interlayer distance
$z$, we use data from DFT+GW calculations \cite{Steinhoff:14, cheiwchanchamnangij_quasiparticle_2012}. 
Assuming a bowing-like behavior for the change of band gap with interlayer distance, a quadratic interpolation leads to
\begin{equation}
E^g_{\mathbf{K}}(z) = 2.675\cdot 10^{-5}z^2 +0.0021z +2.3875
\label{height-bowing}
\end{equation}
for the direct band gap at the $\mathbf{K}$-point.

To introduce the influence of the air gap between the surface and bottom layer within the nanobubble into the TB Hamiltonian, we follow the same route as in the previous section 
and determine the hopping parameter $t_0$ as a function of $E^g_{\mathbf{K}}(z)$. Thereby, the parameter $t_0$ in our TB model depends on the interlayer distance according to.
\begin{equation}
t_0(z) = -2 \left( \frac{E^g_{\mathbf{K}}(z)-(\varepsilon_0 - \varepsilon_2)}{6} + t_{2} + t_{2}' - t_{0}' \right).
\label{t0-bowing}
\end{equation}

\subsection{Normal-normal correlation}
To investigate the wrinkling of the material, we determine the normal-normal correlation function
\begin{equation}\label{eq:nn}
 N_i= \frac{1}{3}\sum\limits_{j,\text{neighbors}} \vec{n}_i\cdot \vec{n}_j
\end{equation}
that measures the angle between the normal vectors of neighboring unit cells. Due to the symmetry of the system, each cell $i$ has three neighboring cells $j$, which gives rise to the 
normalization in equation \eqref{eq:nn}. For better visibility in 
Fig.~\ref{Abb2}b we depict the average angle between adjacent normal vectors $\alpha_i=\arccos(N_i)$.

\subsection{Influence of nanobubble size}
\begin{figure*}[!ht]
\includegraphics[width=0.71\textwidth]{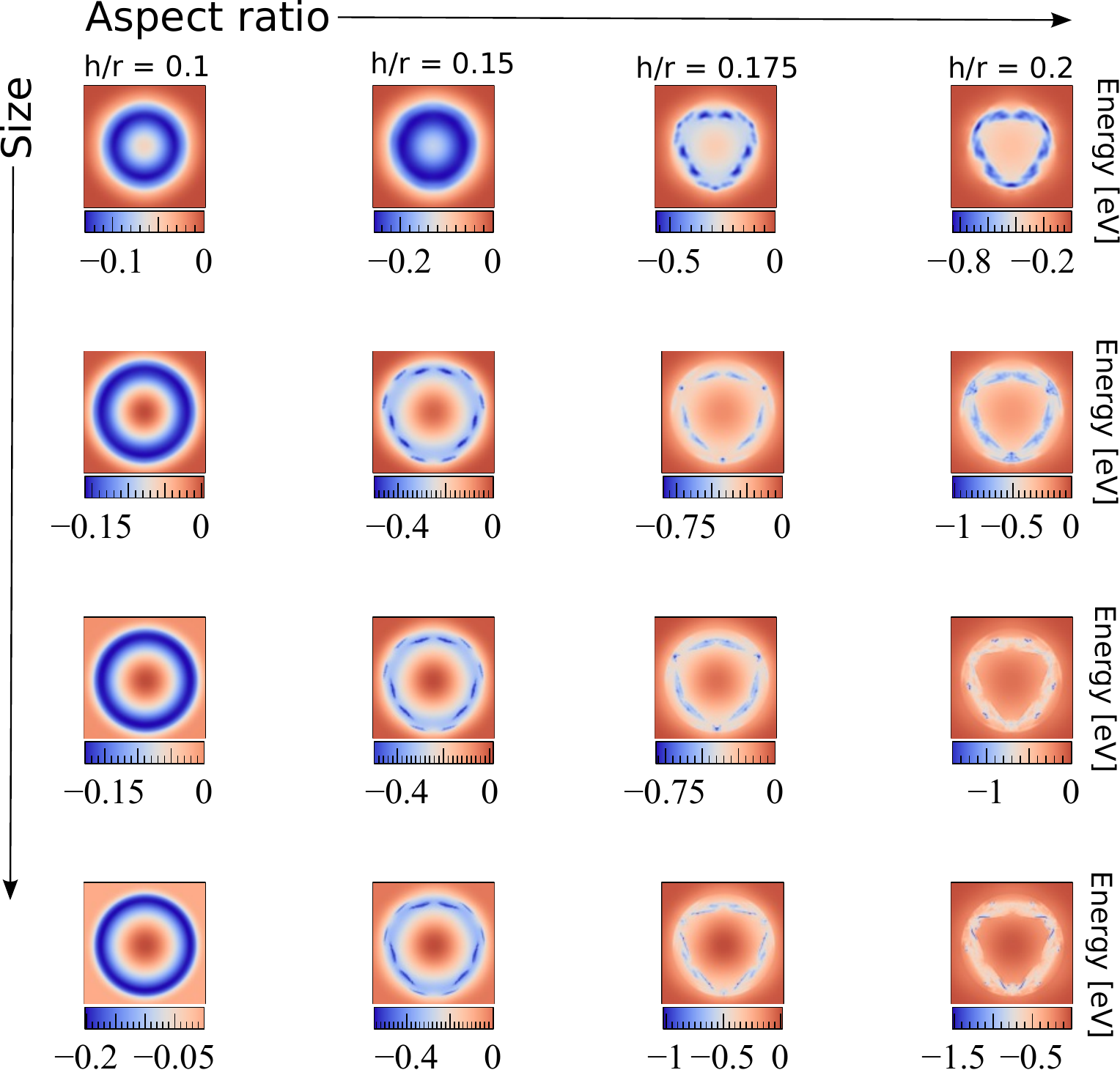}
\caption{Effective confinement potential for aspect ratios of A=0.1,0.15,0.175, and 0.2 and nano-bubble radii of r=6, 12, 18, and 24\,nm.
 \label{Abb_size}}%
\end{figure*}

In Fig.~\ref{Abb_size} we show the influence of both size and aspect ratio variations on the effective confinement potential for the nano-bubble.
We find that the trend towards stronger localization is controlled by the aspect ratio, while the size of the bubble controls the depth of the confinement potential
and the distinctness of the observed localization.
\newpage


\end{document}